\input epsf

\documentstyle[twoside,fleqn,espcrc2,graphicx]{article}

\newcommand{\AmS}{{\protect\the\textfont2
  A\kern-.1667em\lower.5ex\hbox{M}\kern-.125emS}}

\title{Experimental Searches for Non-Baryonic Dark Matter: WIMP Direct Detection\thanks{Invited Review Talk given at the VII Workshop on Topics in Astroparticle and Underground Physics (TAUP2001), Sept.2001, LN Gran Sasso, Italy}}

\author{Angel Morales\thanks{amorales@posta.unizar.es}\\ Laboratory of Nuclear and High Energy Physics and Canfranc Underground Laboratory (LSC) \\
        University of Zaragoza \\ 50009 Zaragoza. Spain}%

\date{}
\begin{document}

% typeset front matter (including abstract)

\begin{abstract}
An overview of the current status of WIMP direct searches is presented, emphasizing strategies, achievements and prospects.
\end{abstract}

\maketitle

\section{Introduction}
Experimental observations and robust theoretical arguments have
established that our universe is essentially non-visible, the
luminous matter scarcely accounting for one per cent of the energy
density of a flat universe. The distribution of a flat universe
($\Omega=\Omega_{M}+\Omega_{\Lambda}=1$) attributes to the dark
energy about $\Omega_{\Lambda}\sim70\%$, whereas the matter
density takes the remaining $\Omega_{M}\sim30\%$, consisting of,
both, visible ($\Omega_{l}\sim0.5\%-1\%$) and non-visible (dark)
matter. This dark component consists of ordinary baryonic matter
($\Omega_{B}\sim 4-5\%$), (possibly made by machos, jupiters,
dust, black holes, etc.) and a large fraction (up to
$\Omega_{NB}\sim25\%$) of non baryonic dark matter, supposedly
made by non-conventional, exotic particles. The minimal
requirements to be fulfilled by the non-baryonic dark particles
are to provide the right relic abundance, to have non-zero mass,
zero electric charge and a weak interaction with ordinary matter.
There are several candidates to such species of matter provided by
schemes beyond the Standard Model of Particle Physics. The
galactic DM axions, the SUSY WIMPs (like neutralinos) and the
light neutrinos (of some non standard models) are particularly
attractive. The WIMPs, Weak Interacting Massive Particles, are
favorite of experimentalists ($\sim20$ experiments for WIMPs vs. 2
for axions) and so of this talk. This talk deal with the recent
efforts done in its direct search illustrated by a few, selected
experiments.

\begin{table*}[htb]
\caption{WIMP Direct Detection in underground facilities
experiments currently running (or in preparation) }
\label{table:2}
\begin{tabular}{lll}
\hline LABORATORY &  EXPERIMENT & TECHNIQUE \\ \hline

BAKSAN (Russia) & IGEX &  3$\times$1 Kg Ge-ionization \\

BERN(Switzerland) &  ORPHEUS &(SSD) Superconducting Superheated
Detector, 0.45 Kg Tin \\

BOULBY & NaI & NaI scintillators of few Kg (recently completed) \\
(UK)& NAIAD &  NaI unencapsulated scintillators (50 Kg)\\

& ZEPLIN & Liquid-Gas Xe scintillation/ionization  I: 4 Kg single
phase    \\ & & II: 30 Kg Two phases \\

&    DRIFT &  Low pressure Xe TPC (in preparation) 1
m$^{3}\rightarrow$ 10 m$^{3}$ \\

CANFRANC  &  COSME  & 234 g Ge ionization  \\

(Spain) & IGEX  &  2.1 Kg Ge ionization \\

&    ANAIS &  10$\times$10.7 Kg NaI scintillators \\

&    ROSEBUD & 50g Al$_{2}$O$_{3}$ and 67g Ge thermal detectors \\
& &        CaWO$_{4}$ 54g and BGO 46g scintillating bolometers \\

FREJUS/MODANE &  SACLAY-NaI &   9.7 Kg NaI scintillator (recently
completed) \\

(France) &   EDELWEISS I &70 g Ge thermal+ionization detector \\
 &   EDELWEISS II  &  4$\times$320 g Ge thermal+ionization
 detectors \\

GRAN SASSO  &H/M &2.7 Kg Ge ionization \\

 (Italy) &HDMS   & 200g Ge ionization in Ge well \\

&    GENIUS-TF  & 40$\times$2.5 Kg unencapsulated Ge (in
preparation) \\

&    DAMA   & NaI scintillators (87.3 Kg) \\

&    LIBRA  & NaI scintillators 250 Kg (in preparation) \\

& Liquid-Xe &  Liquid Xe scintillator (6 Kg) \\

&    CaF$_{2}$ &   Scintillator \\

&    CRESST I &   (4$\times$260g) Al$_{2}$O$_{3}$ thermal
detectors \\

&    CRESST II  & Set of 300g CaWO$_{4}$ scintillating bolometers
(up to 10 Kg) \\

&    MIBETA  & 20$\times$340g TeO$_{2}$ thermal detector \\

&    CUORICINO &  56$\times$760g TeO$_{2}$ thermal detector (being
mounted) \\

&    CUORE   &1000$\times$760g TeO$_{2}$ (in preparation) \\

RUSTREL (France)&   SIMPLE & (SDD)Superheated Droplets Detectors
(Freon)
\\

STANFORD UF/ & CDMS - I &   100g Si; 6$\times$165g Ge
thermal+ionization detectors \\

SOUDAN(USA)&   CDMS - II &  3$\times$250g Ge and 3$\times$100g Si
Thermal+Ionization \\

SNO (Canada) & PICASSO &(SDD)Superheated Droplets Detectors (1.34g
of Freon) \\

OTO & ELEGANTS-V &   Large set of massive NaI scintillators
\\
 (Japan) &ELEGANTS-VI &  CaF$_{2}$ scintillators \\ \hline

\end{tabular}\\[2pt]
\end{table*}

Galactic halo WIMPs could be directly detected by measuring the
nuclear recoil produced by their elastic scattering off target
nuclei in suitable detectors at a rate which depends of the type
of WIMP and interaction. In the case of WIMPs of $m\sim
GeV~\mathrm{to}~\textit{TeV}$ and $v\sim10^{-3}c$ the nuclear
recoil in the laboratory frame
$E_{R}=\frac{\mu^{2}}{M}v^{2}(1-cos\theta)$ is in the range from 1
to 100 \textit{KeV}. $M$ is the nuclear mass, $\mu$ the ($m, M$)
reduced mass and $\theta$ the WIMP-nucleus (c. of m.) scattering
angle. Only a fraction $QE_{R}=E_{vis} (\equiv E_{eee})$ of the
recoil energy is visible in the detector, depending on the type of
detector and target and on the mechanism of energy deposition. The
so-called Quenching Factor Q is essentially unit in thermal
detectors whereas for the nuclei used in conventional detectors it
ranges from about 0.1 to 0.6. The energy delivered by the WIMP
results in a small signal (1-100 KeV) which shows up even smaller
(for Q$<$1). Moreover this signal falls in the low energy region
of the spectrum, where the radioactive and environmental
background accumulate at much faster rate and with similar shape.
That makes WIMP signal and background practically
undistinguishable. On the other hand, the smallness of the
neutralino-matter interaction cross-section implies that the
process looked for is very rare.

Customarily, one compares the predicted event rate with the observed spectrum. If the former turns out to be larger than the
measured one, the particle which would produce such event rate can be ruled out as a Dark Matter candidate. That is expressed as
a contour line $\sigma$(m) in the plane of the WIMP-nucleon elastic scattering cross section versus the WIMP mass. That excludes,
for each mass m, those particles with a cross-section above the contour line $\sigma$(m). The level of background sets,
consequently, the sensitivity of the experiment in eliminating candidates or in constraining their masses and cross sections.

However, this simple comparison will not be able to identify the
WIMP. A convincing proof of its detection should be provided by a
distinctive signature characteristic of WIMPs. Such distinctive
labels do exist: they are originated by the motion of the Earth in
the galactic halo \cite{Drukier,Spergel}. These signatures are an
annual modulation of the rate and a directional asymmetry of the
nuclear recoil. Narrowing  first the window of the possible WIMP
existence and looking then for its identification is the purpose
of the experimental searches. Table \ref{table:2} gives an
overview of the experiments on direct detection of WIMPs currently
in operation or in preparation, which is the subject of this talk.
General reviews for WIMP dark matter are given in Ref. \cite{Gri}.
WIMP direct detection is reviewed, for instance in Ref.
\cite{Mor2,Mor}.

%\begin{figure*}[ht]
%\centerline{\includegraphics[height=18cm]{wddufecr.ps} } Table 1
%\end{figure*}

WIMPs can be also looked for, indirectly, in the galactic halo,
looking for its presence in cosmic ray experiments in terms of
antiprotons, positrons or gamma rays produced by WIMP annihilation
in the halo. One can also search in underground, underwater or
under-ice detectors, looking (also indirectly) for WIMPs through
the high energy neutrinos emerging as final products of the WIMP
annihilation in celestial bodies (Earth or Sun).

\section{Strategies for WIMP direct detection}

In order to plan an experiment it is essential to know how small
the WIMP-nucleus interaction rates are, to estimate which are the
chances of a positive result, for a given experimental
sensitivity. According to the SUSY model employed and the choice
of the various parameters, the predicted rate encompass various
orders of magnitude, going from $R\sim1-10~c/Kg day$ down to
$10^{-4}-10^{-5}~c/Kg day$. The $\sigma_{\chi N}$ calculations are
made within the Minimal Supersymmetric extension of the Standard
Model, MSSM, as basic frame, implemented in various schemes (see
Ref.\cite{treinta}). Besides the peculiarities of the SUSY model
there is a wide choice of parameters entering in the calculation
of the rates: the halo model, the values of the WIMP velocity
distribution parameters, the three levels of the WIMP-nucleus
interaction (quark-nucleon-nucleus) and the constraint of getting
the proper relic abundance of the candidates. The theoretical
prediction of the rates are presented as "scatter plots" extending
along the various orders of magnitude quoted above. Some of the
most favourable predictions are already testable by the leading
experiments which have, in fact, penetrate into the scatter plot
of predictions. The bottom of the plot is still far away of the
detector sensitivity. It should be noted, however, that most of
the experimental searches for SUSY-WIMP's concentrate in the
dominant, coherent interaction, which provides the largest
signals.

How the WIMP direct search proceeds? To face the challenge of
detecting signals of small size ($E_{R}<100~KeV$) produced at a
very small rate ($R\sim10\rightarrow10^{-5}~c/Kg day$) (equivalent
to detecting processes of $\sigma\sim10^{-10}~pb$ (for a Ge
detector)), which have a similar shape than the background,
ultra-low background detectors and a radioactivity-free
environment are needed. Example of low background recently
achieved is the case of IGEX, which in its raw Ge background
spectra have reached a level of $10^{-1}\sim10^{-2}~ c/KeV Kg day$
in the low energy region, and the cases of the CDMS and EDELWEISS
experiments where their present total rate is of $\sim1~c/Kg day$.
Because of the small recoil energy delivered in the WIMP
interaction, detectors of very low energy threshold and high
efficiency are mandatory. That is the case for the bolometer
experiments (MIBETA, CRESST, ROSEBUD, CUORICINO, CDMS and
EDELWEISS) which seen efficiently the energy delivered by the WIMP
(quenching factor is unity) and which have achieved very low
energy thresholds ($E_{THR}\sim$ hundreds of $eV$). Large masses
of targets are also recommended, to increase the probability of
detection and the statistics. The DAMA, UKDMC and Zaragoza
scintillations experiments use from 50 to 100 Kg of NaI; the
CUORICINO bolometer experiment plans to employ 42 Kg of $TeO_{2}$
crystals and the ZEPLIN detector which uses (according to the
various versions) from 20 to 30 Kg of Xe. It is remarkable that
small size, first generation detectors have reached exclusions
$\sigma_{\chi p}>10^{-5}-10^{-6}~pb~(10^{-41}-10^{-42}~cm^{2})$ in
the range of masses relevant for SUSY-WIMPs.

To get an ultralow background, the radiopurity of the detector,
components, shielding and environment must be the best achievable
with the current state-of-the-art. The next step is to reject
components of the background by discriminating electron recoils
(tracers of the background) from nuclear recoils (originated by
WIMPs and neutrons). Methods used to discriminate backgrounds from
nuclear recoils are either simply statistical, like a Pulse Shape
analysis (PSD), based on the different timing behaviour of both
types of pulses, or event by event by measuring simultaneously,
the ionization (or scintillation) and the heat, and noting that
for a given deposited energy (measured as phonons) the recoiling
nucleus ionizes less than the electrons. Examples of PSD are the
sodium iodide experiments of UKDMC, Saclay, DAMA and ANAIS. Event
by event discrimination has been successfully applied in CDMS and
EDELWEISS by measuring ionization and heat and now in CRESST and
ROSEBUD by measuring light and heat.

Another discriminating technique is that used in the two-phase liquid-gas Xenon detector with ionization plus scintillation, of
the ZEPLIN series of detectors. An electric field prevents recombination, the charge being drifted to create a second pulse in
addition to the primary pulse. The amplitudes of both pulses are different for nuclear recoils and electrons allowing their
discrimination.

One could use instead threshold detectors--like neutron
dosimeters--which are blind to most of the low Linear Energy
Transfer (LET) radiation (e, $\mu$, $\gamma$) and so able to
discriminate gamma background from neutrons (and so WIMPs).
Detectors which use superheated droplets which vaporize into
bubbles by the WIMP (or other high LET particles) energy
deposition are those of the SIMPLE and PICASSO experiments. An
ultimate discrimination will be the identification of the
different kind of particles by the tracking they left in, say, a
TPC, plus the identification of the WIMP through the directional
sensitivity of the device (DRIFT).

\section{Excluding WIMPs which cannot be dark matter}

How to proceed to exclude a WIMP candidate? The direct experiments
measure the differential event rate (energy spectrum) in the
customary differential rate unit (dru)
$\left[\frac{dR}{dE_{\mathrm{VIS}}}\right]^{\mathrm{exp}}(c/KeV Kg
day)$. The registered counts R contain the signal and the
background. Then, by applying discrimination techniques one
disentangle at least partially the nuclear recoils from the
background events. The resulting residual rate is then compared
(in terms of m, and $\sigma_{\chi N}$) with the theoretical
nuclear recoil rate due to WIMPs interaction (in counts per KeV Kg
day)
$$\left[\frac{dR}{dE_{\mathrm{VIS}}}\right]^{\mathrm{Th}}=7.76\times10^{14}\frac{N}{Q}\frac{\rho}{v_{E}}\frac{(m+M)^{2}}{4m^{3}M}F^{2}\sigma^{SI}_{\chi
N}\tau$$ (with $\sigma$ in $cm^{2}$, m and M in GeVs, $v$ in
$Kms^{-1}$, $\rho$ in $GeVcm^{-3}$ and N in $Kg^{-1}$) where
$\rho$ is the local density of WIMP, N the number of target
nuclei, $F^{2}$ is the nuclear form factor, and
$\tau(v_{\mathrm{esc}}\rightarrow\infty)=erf(x+y)-erf(x-y)$, with
$$x,y=\sqrt{\frac{3}{2}}(v_{\mathrm{min}},v_{E})\frac{1}{v_{\mathrm{rms}}}~~,~~~~v_{\mathrm{rms}}\sim\sqrt{\frac{3}{2}}v_{\mathrm{sun}}$$
$v$ the WIMP velocity (Earth/Lab frame) and $v_{E}$ the velocity
of Earth/Solar system with respect to the halo.

$v_{min}(E_{R})=\frac{m+M}{m}(E_{R}/2m)^{\frac{1}{2}}$ is the minimal velocity to produce a recoil $E_{R}$. The spin-independent
nuclear cross-section is usually normalized in terms of that on nucleons
$$\sigma^{SI}_{\chi N}=A^{2}\frac{\mu^{2}_{\chi N}}{\mu^{2}_{\chi n}}\sigma^{\mathrm{scalar}}_{\mathrm{nucleon(p.n)}}$$

Those values of ($\sigma$, m) predicting a recoil spectrum above the observed rate
$$\left[\frac{dR}{dE_{\mathrm{VIS}}}\right]^{\mathrm{Th}}\geq\left[\frac{dR}{dE_{\mathrm{VIS}}}\right]^{\mathrm{exp}}_{\mathrm{UppBound}}$$ are
excluded. The region above the contour $\sigma(m)$ is depicted as an exclusion plot of those WIMPs of mass $m$ with interaction
cross-section above $\sigma$. Obviously, the smaller the background the better the exclusion. However excluding a WIMP is not
enough.
%$$\sigma^{SI}_{\chi N}=A^{2}\frac{\mu^{2}_{\chi N}}{\mu^{2}_{\chiP}}\sigma^{\mathrm{scal}}_{\mathrm{p}}$$

\section{The identification of WIMP dark matter}

After reducing maximally the background and extremating the
discrimination (99.99 \%) of the detector, one should look for
asymmetries characteristic of WIMP signals. Typical smoking guns
of WIMPs would be the annual modulation of the rate
\cite{Drukier}, the forward/backward asymmetry of the nuclear
recoil \cite{Spergel} or the nuclear target dependence of the
rates \cite{Smith}.

The two kinematical asymmetries characteristic of WIMPs signals
are originated by the Earth motion through the galactic halo. The
Earth orbital motion around the Sun has a summer/winter variation,
which produces a small annual modulation of the WIMP interaction
rates, of the order
$O\left(\frac{v_{\mathrm{rot,E}}}{v_{h}}\right)\sim\frac{15}{270}\sim5\%$
\cite{Drukier}. The observation of this small modulation of a very
small signal requires large target mass and exposure, superbe
stability and extreme control of systematics and of other
stational effects.

The orbital velocity of Earth around the sun is of 30 $Kms^{-1}$
in an orbit inclined $\alpha=60^{\circ}$ with respect to the
galactic disk
$$v_{E,r}=30\cos\alpha\cos\omega(t-t_{0})\rightarrow15\cos\omega(t-t_{0})Kms^{-1}$$
$$\omega=\frac{2\pi}{T}~~~~~T=1 \mathrm{year}~~~~t_{0}:
\mathrm{June}~2^{nd}$$ So, the velocity of the Earth (and of our
earthborne detector) relative to the galactic halo is
$$v_{E}=v_{\mathrm{sun}}+15\cos\omega(t-t_{0})Kms^{-1}$$
Consequently, in summer there is a component of the Earth' motion
around the sun parallel to the sun motion through the galaxy which
adds 15 $Kms^{-1}$. On the contrary, in winter the same occurs but
the motion is antiparallel and so one has to subtract 15
$Kms^{-1}$. The result is that the detector moves slightly faster
in June than in December (5\% effect), and consequently a
modulation of the WIMP interaction rates follows, given at first
order by $$S(t)=S_{0}+S_{m}\cos\omega(t-t_{0})[+B]$$ where $S_{0}$
is the average signal amplitude, $S_{m}$ the modulated amplitude
and $B$ the constant background.
%This effect has been explored in the
%experiment NaI-32 (Zaragoza), using 32 Kg of NaI in Canfranc; in the LqXe (Roma) in Gran Sasso; in the DEMOS experiment
%(USC-PNL-Zaragoza-Tandar), with a Ge detector in Sierra Grande and in the ELEGANTS (Osaka) experiment with a large amount of NaI
%scintillators in Kamioka. A positive result of such annual variation has been reported by the DAMA NaI (Roma) experiment an
%atributed by the Collaboration to a WIMP of $m\sim60 GeV$ and scalar cross-section on protons of $\sigma\sim 7\times
%10^{-6}picobarns$.

A second characteristic signature of the WIMP is provided by the
directional asymmetry of the recoiling nucleus \cite{Spergel}. The
WIMPs velocity distribution in the Earth frame is peaked in the
opposite direction of the Earth/Sun motion through the halo, and
so the distribution of nuclear recoils direction shows a large
asymmetry forward/backward (F/B) not easily mimicked by the
supposedly isotropic background. The order of magnitude of the
effect is large because the solar system' motion around the
galactic center $v_{\mathrm{sun}}$, and the typical WIMP velocity
in the halo, $v_{h}$, are of the same order
$O\left(\frac{v_{\mathrm{sun}}}{v_{h}}\right)\sim\frac{230}{270}\sim1$.

The angular dependence of event rate is given by (see
Ref.\cite{Spergel}) $$\frac{d^{2}R}{dE_{R}d(\cos\gamma)}=$$
$$\frac{N\rho\sigma}{\sqrt{\pi}}\frac{(m+M)^{2}}{2m^{3}Mv_{\mathrm{halo}}}~~\mathrm{exp}\left\{-\frac{(\textit{v}_{E}\cos\gamma-\textit{v}_{\mathrm{min}})^{2}}{\textit{v}^{2}_{h}}\right\}$$
where $\gamma$ is the angle of the recoiling nucleus.

%The F/B asymmetry is a clear WIMP signature not easily mimicked by the supposedly isotropic background. It is larger and more
%clearly distinguishable from the background than the annual modulation. Various forthcoming projects with sensitivity to
%directionality (TPC, DRIFT) in different status of R+D, have triggered a large activity in analyzing
Recently, the nuclear recoil angular dependence of WIMP
interactions has been analyzed in different halo models with the
purpose of exploring how well a directional signal can be
distinguished without ambiguity from the background -with
independence of the halo model-. Quite remarkably if the device
has angular resolution sensitivity, few events will be enough to
distinguish the signal, and not too many are needed if it has only
F/B sensitivity. See Ref.\cite{Copi}.

Another asymmetry is the nuclear target dependence of the rate
\cite{Smith}. However, due to the differences in the intrinsic
backgrounds of the various targets, it is not easy to get reliable
conclusions. Some experiments are operating (or can do it) sets of
similarly produced crystals of different nuclear targets in the
same environment like ROSEBUD ($Ge/Al_{2}O_{3}/CaWO_{4}$).

\begin{figure}[t]
\centerline{\includegraphics[height=5.2cm]{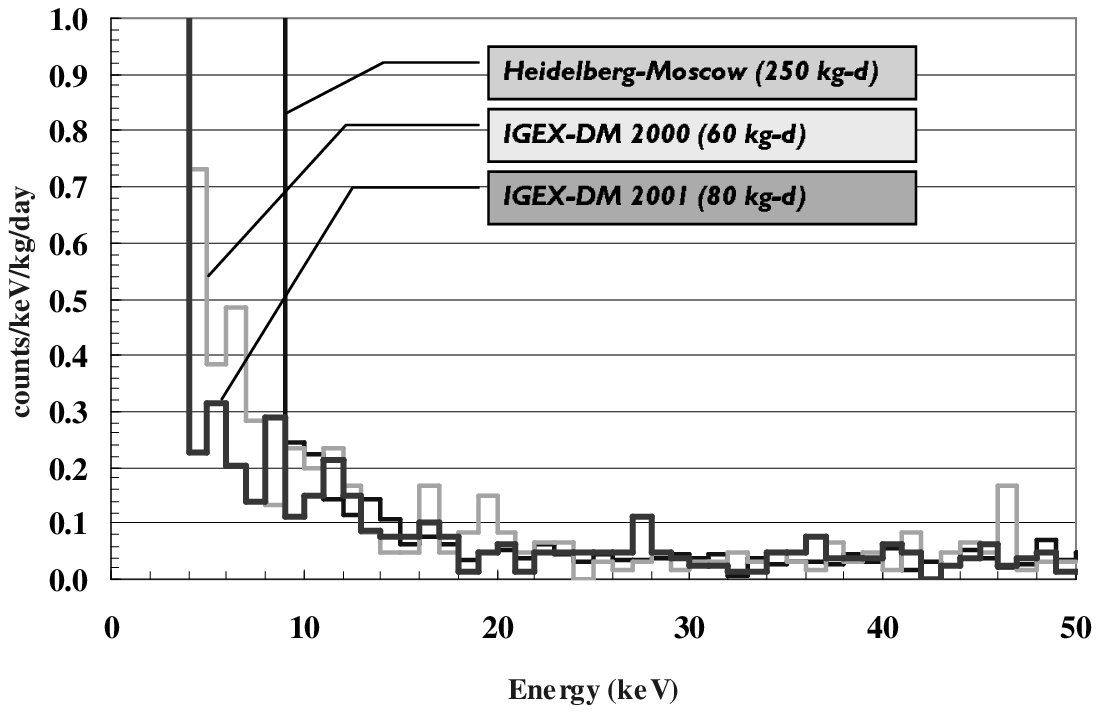}} \caption{}
\label{fig1}
\end{figure}

\section{Germanium Experiments}

The high radiopurity and low background reached in Germanium
detectors, the low energy threshold which can be obtained, their
reasonable quenching factor (about 25\%) and other nuclear merits
make Germanium a good option to search for WIMPs with detectors
and techniques fully mastered \cite{sitges}.
%Table 2 shows the Germanium ionization detector experiments currently in operation, and their main
%features. The best results have been obtained recently by the IGEX experiment.

The International Germanium Experiment (IGEX) \cite{Aal} is using
one enriched detector of $^{76}$Ge of $\sim$2.1 Kg to look for
WIMPs in the Canfranc Underground Laboratory. It has an energy
threshold of 4 keV and an energy resolution of 0.8 keV at the
75~keV Pb x-ray line. The detector is fitted in a cubic block in
lead being surrounded by not less than 40-45 cm of lead of which
the innermost 25 cm are archaeological. A muon veto and a neutron
shielding of 40 cm of polyethylene and borated water completed the
set-up \cite{Mor00,Mor01}. The spectrum of IGEX-2001 \cite{Mor01}
together with that of a previous run \cite{Mor00} (IGEX-2000) are
shown in Fig. \ref{fig1} in comparison with that of the
Heidelberg-Moscow experiment. The H/M experiment\cite{Bau99} is
another enriched-Ge experiment (energy threshold of 9 KeV),
already completed, which has been running at Gran Sasso.

%\begin{figure}[htb]
%\centerline{ \fboxrule=0cm
% \fboxsep=0cm
%  \fbox{
%\epsfxsize=5cm
%  \epsfbox{fig1.eps}}}
%\begin{center}
%{\caption {}\label{fig1}}
%\end{center}
%\end{figure}

%The method followed in deriving the plots has been the same for IGEX and H/M. As recommended by the Particle Data Group, the
%predicted signal in an energy bin is required to be less than or equal to the (90\% C.L.) upper limit of the (Poisson) recorded
%counts. The derivation of the interaction rate signal supposes that the WIMPs form an isotropic, isothermal, non-rotating halo of
%density $\rho = 0.3$~GeV/cm$^{3}$, have a Maxwellian velocity distribution with $\rm v_{\rm rms}=270$~km/s (with an upper cut
%corresponding to an escape velocity of 650~km/s), and have a relative Earth-halo velocity of $\rm v_{\rm r}=230$~km/s. The cross
%sections are normalized to the nucleon, assuming a dominant scalar interaction. The Helm parameterization\cite{Eng91} is used for
%the scalar nucleon form factor, and the recoil energy dependent ionization yield used is the same that used by the H/M
%collaboration in Ref \cite{Bau99} $\rm E_{vis} = 0.14 (E_{REC}) ^{1.19}$.

\begin{figure}[t]
\centerline{\includegraphics[height=6cm]{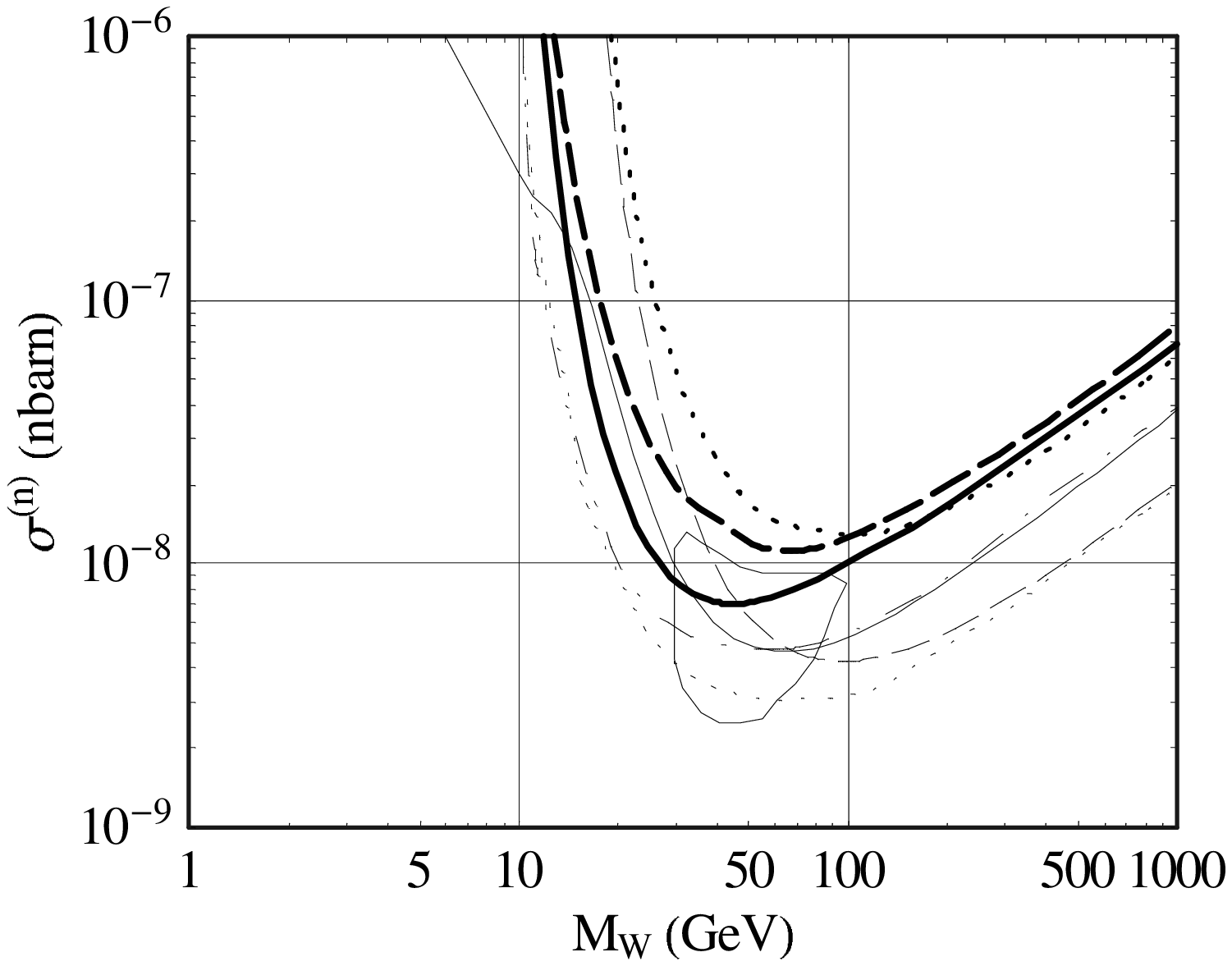}} \caption{}
\label{fig2}
\end{figure}

The exclusion plots derived for these three Ge experiments are
depicted in Fig. \ref{fig2} following the same method: IGEX-2001
(thick solid line), IGEX-2000 (thick dashed line) and
Heidelberg-Moscow (thick dotted line). IGEX-2001 improves the
exclusion of all other Ge-ionization experiments for a mass range
from 20~GeV up to 200~GeV, which encompass the DAMA mass region
\cite{Ber99}. In particular, this new IGEX result excludes
WIMP-nucleon cross-sections above 7 $\times 10^{-6}$ pb for masses
of 40-60 GeV and enters the DAMA region excluding the upper left
part of this region. That is the first time that a direct search
experiment with a Ge-diode without background discrimination, but
with very low (raw) background, enters such region. A further 50
\% background reduction between 4 keV and 10 keV (which could be
reasonably expected) would allow IGEX to explore practically all
the DAMA region in 1~kg~y of exposure. (We refer to the Igor G.
Irastorza contribution to these Proceedings and
Ref.\cite{sitges,Mor01}).

Also shown for comparison are the contour lines of the other
experiments, CDMS\cite{Abusaidi:2000} and
EDELWEISS\cite{Benoit:2001} (thin dashed line), which have entered
that region by using bolometers which also measure ionization and
which will be described later on. The DAMA region (closed line)
corresponding to the annual modulation effect reported by that
experiment \cite{Ber99} and the exclusion plot obtained by DAMA
NaI-0 \cite{trece} (thin solid line) by using statistical pulse
shape discrimination are also shown. In the CDMS case two contour
lines have been depicted, the exclusion plot published in
Ref.\cite{Abusaidi:2000} (thin dotted line) and the CDMS expected
sensitivity contour (thin dot-dashed line) (See B. Sadoulet
contribution to these Proceedings.)

%\begin{figure}[ht]
%\centerline{\includegraphics[height=5.4cm]{ExclusionIGEX.eps}}
% \caption{IGEX-DM 2001 exclusion plot for spin-independent
%interaction obtained in this work (thick solid line) compared with the previous exclusion obtained by IGEX-DM 2000 (dashed thick
%line) and the last result obtained by the Heidelberg-Moscow germanium experiment cite{Bau} (dotted line) recalculated from the
%original spectrum with the same set of hypothesis and parameters. The closed line corresponds to the (3$\sigma$) annual
%modulation effect reported by the DAMA collaboration (including NaI-1,2,3,4 runnings). The thin solid line is the exclusion line
%obtained by DAMA NaI-0 by using Pulse Shape Discrimination. The two other experiments which have entered the DAMA region are also
%shown: EDELWEISS (thin dashed line) and the CDMS exclusion contour (thin dotted line) and its expected sensitivity (thin
%dot-dashed line).
%%Also
%%shown are the other three experiments which enters in the DAMA
%%region: CDMS \cite{Abusaidi:2000} (dot-dashed thin line),
%%EDELWEISS \cite{Benoit:2001} (dashed thin line) and DAMA NaI-0
%%\cite{Ber96} (thin solid line).
%} \label{dm-ig-5}
%\end{figure}

%\begin{figure*}
% \centerline{\includegraphics[height=10cm]{geioex.ps}}
%\end {figure*}

%\begin{figure*}
% \centerline{\includegraphics[height=10cm,angle=-90]{geioex.eps}}
%\end {figure*}

%Data collection is currently in progress in the IGEX experiment and
%some strategies are being considered to further reduce the
%low energy background.

A new experiment GEDEON (GErmanium DEtectors in ONe cryostat) is
planned for Canfranc \cite{sitges}. It will use the technology
developed for the IGEX experiment. The cell-unit will consist of a
set of $\sim$1 kg germanium crystals, of a total mass of about 28
kg, placed together in a compact structure inside one only
cryostat. Expected backgrounds are $\sim 10^{-3}~c/KeV Kg day$ in
the low energy region. A total mass of $\sim 112$ Kg of Ge (or 4
cells) is projected for a second phase of GEDEON.

%\begin{figure}[ht]
%\centerline{\includegraphics[height=5.4cm]{ProspectsIGEX.eps} }
% \caption{IGEX-DM projections are shown for
% a flat background rate of 0.1~c/keV/kg/day (dot-dashed line) and 0.04~c/keV/kg/day (solid line) down
% to the threshold at 4 keV, for 1~kg~year of exposure.
% The exclusion contour expected for GEDEON is also
% shown (dashed line) as explained in the text.} \label{dm-ig-6}
%\end{figure}

Two more Ge experiments running or in preparation, both in Gran
Sasso, are that of the Heidelberg Dark Matter Search (HDMS) and
the GENIUS-Test Facility. The small detector of HDMS has achieved
a background still higher than that of H/M and so the results will
not be include here. See Ref.\cite{bau00}. Most of the attention
of this Collaboration goes now to the preparation of a small
version (GENIUS-Test Facility) of the GENIUS project
\cite{genius}. GENIUS is a multipurpose detector, consisting of
enriched Ge detectors of about 2$\frac{1}{2}$ Kg each (up to a
total of 0.1 to 10 tons) which uses the novel idea of immerse the
crystals directly into a large tank of liquid nitrogen. GENIUS-TF
(fourteen detectors of natural isotopic abundance and low nominal
energy threshold), now in preparation, is intended to test the
GENIUS project and at the same time to search for WIMP. (We refer
to the H.V. Klapdor contribution to these Proceedings for
details).

\section{WIMP searches with NaI scintillators}

The sodium iodide detectors are very attractive devices to look
for WIMP. Both nuclei have non-zero spin
($^{23}Na~J=\frac{3}{2},~^{127}I~J=\frac{5}{2}$) and then
sensitive also to spin dependent interaction. Iodine is a heavy
nucleus favourable for spin-independent interactions. The
quenching factor is small ($Q<10\%$) for I, and medium for Na
($Q\sim30-40\%$). Backgrounds lesser than or of the order of $\sim
1$ count per KeV Kg day in the few KeV region have been achieved.
There exists four NaI experiments running: DAMA, UKDMC, (in
various detectors and projects), ELEGANTS and ANAIS.

\begin{figure}[t]
\centerline{\includegraphics[height=5.4cm]{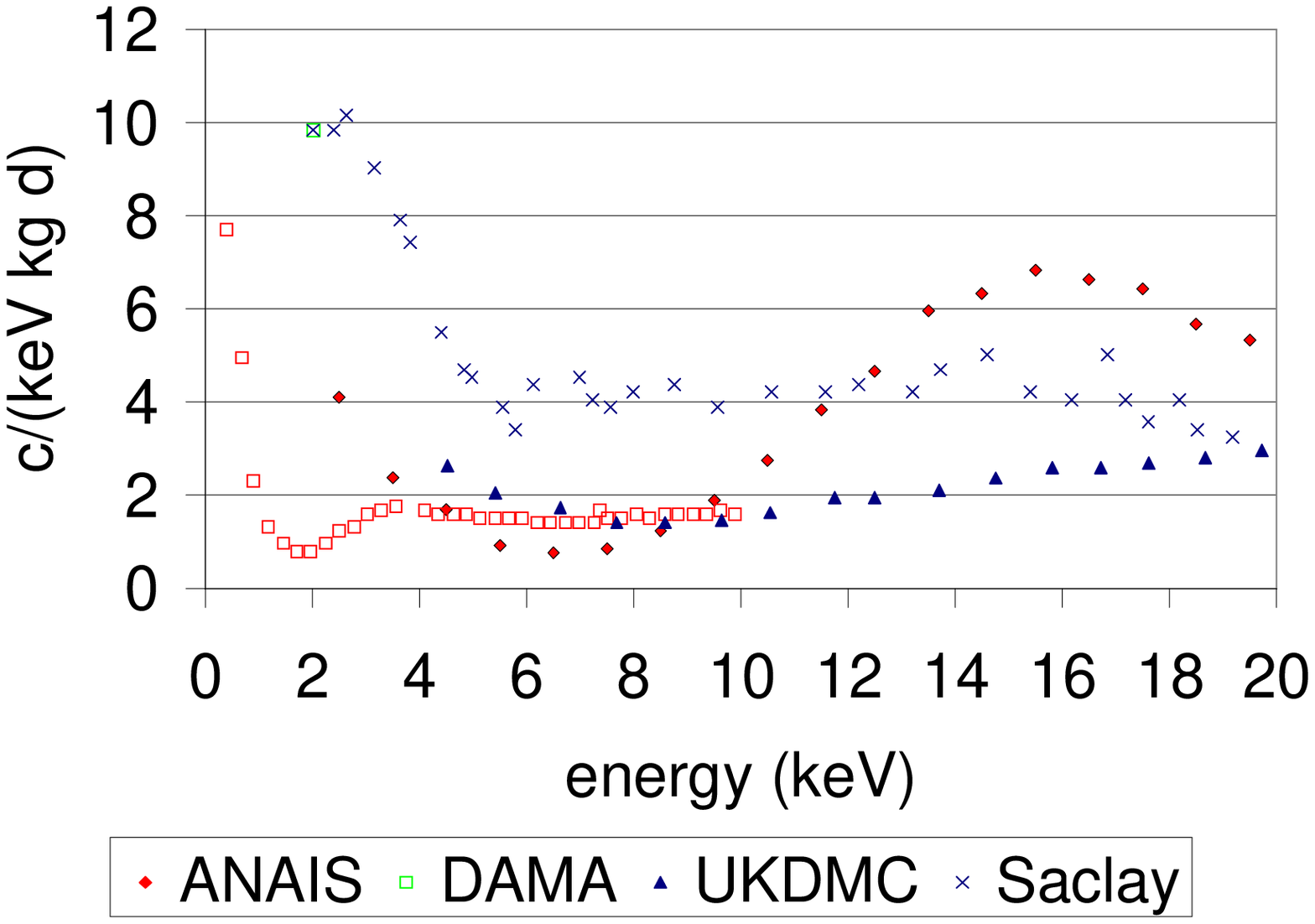}} \caption{}
\label{fig3}
\end{figure}

The NaI scintillators can be endowed with Pulse Shape
Discrimination (PSD) to distinguish statistically gamma background
from WIMPs (or neutron) signals, because of the different timing
behaviour of their pulses. From such statistical analysis it
results that only a few percent (depending on the energy) of the
measured background can be due to nuclear recoils. The background
spectra (before PSD) of the four NaI experiments ANAIS
\cite{cebrian}, DAMA \cite{Ber99}, UKDMC \cite{ukdm} and Saclay
\cite{saclay} are shown comparatively in Fig. \ref{fig3} (1 to 2
$c/KeV Kg day$ in DAMA, UKDMC and ANAIS and of 2 to 10 $c/KeV Kg
day$ in Saclay and ELEGANTS \cite{elegants}).

The United Kingdom Dark Matter Collaboration (UKDMC) uses
radiopure NaI crystals of various masses (2 to 10 kg) in various
shielding conditions (water, lead, copper) in Boulby \cite{ukdm}.
Typical thresholds of 4 keV have been obtained. Results from NaI
crystals of various dimension showed a small population
("anomalous") of pulses of an average time constant shorter than
that of gamma events and attributed probably to an alpha surface
contamination. New crystals with polished surfaces (and operating
unencapsulated in dry nitrogen atmosphere) do not show such
anomaly. UKDMC is also preparing, NAIAD (NaI Advanced Detector)
wich will consist of 50--100 kg in a set of unencapsulated
crystals to avoid surface problems and improve light collection.
(See S. Hart contribution to these Proceedings.)

\begin{figure}[t]
\centerline{
\includegraphics[height=4.0cm]{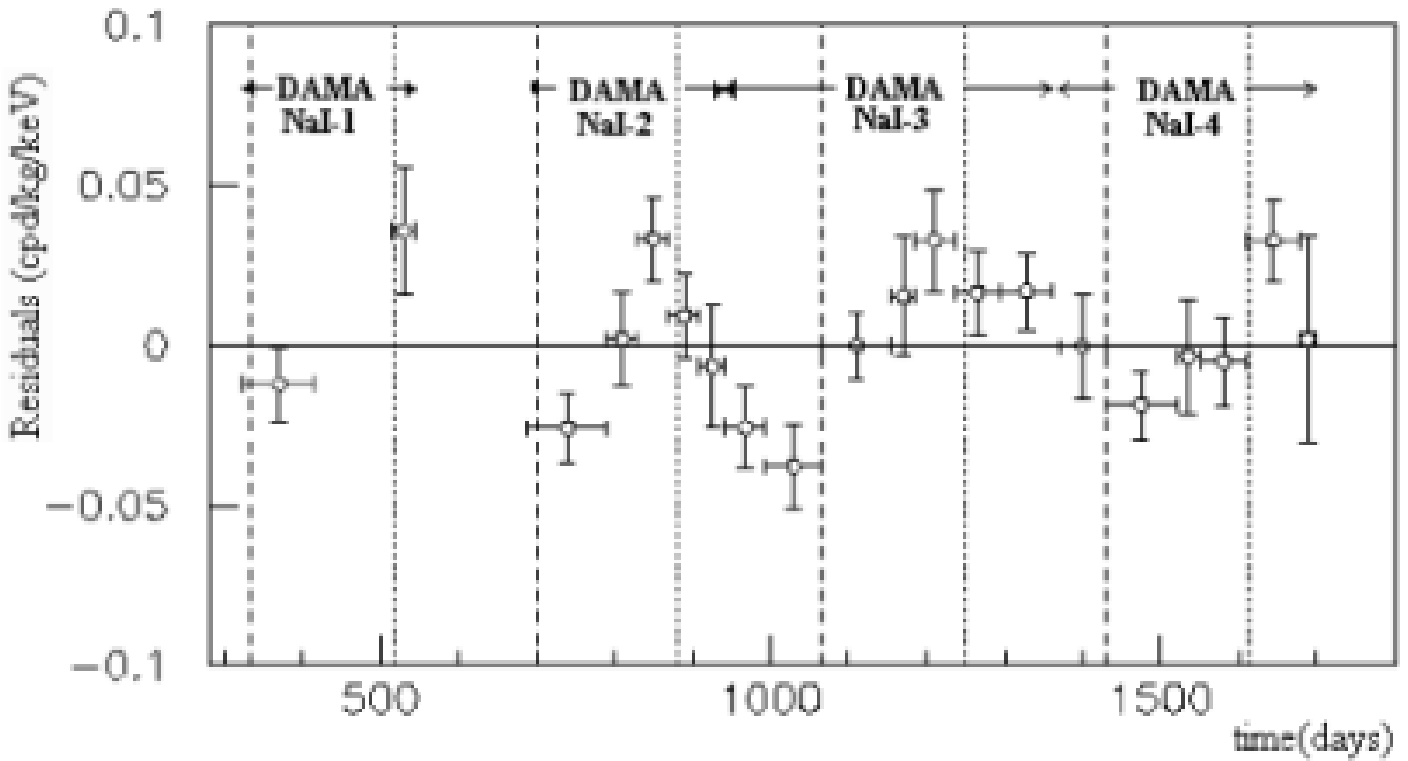}} \caption{} \label{fig4}
\end{figure}

ANAIS (Annual Modulation with NaI's) will use 107 kg of NaI(Tl) in
Canfranc. A prototype of one single crystal (10.7 kg) is being
developed. The components of the photomultiplier have been
selected for its radiopurity. Pulse Shape analysis has been
performed and the possible presence of "anomalous" events
investigated. No evidence of such anomaly has been found. The
preliminary results refer to an exposure of 1225.4 $kg day$. Fig.
\ref{fig3} shows the background after the noise rejection. The
energy threshold is of $\sim 4$ keV and the background level
registered from threshold up to 10~keV is about 1.2 $c/KeV Kg
day$. (We refer to the S. Cebri\'{a}n contribution to these
Proceedings).

The DAMA experiment \cite{Ber99} uses 9 radiopure NaI crystals of
9.7 Kg each, viewed by two PMT. The background spectrum is shown
also in Fig. \ref{fig3}. The software energy threshold is at
$E_{Thr}=2keV$ and the energy resolution at 2-5 keV is
$\Gamma\sim2-2.5 keV$. The PSD method applied to the DAMA NaI-0
\cite{trece} running lead to a background reduction of 85\% (at
4--6 keV) and 97\% (at 12--20 keV), providing exclusion plots
which have surpassed that of germanium.

The main objective of DAMA is to search for the annual modulation
of the WIMPs signal. Such modulation has been found and attributed
by the Collaboration to a WIMP signal. After 57986 Kg day of
statistics the residuals of the rate vs time, looks as shown in
Fig.\ref{fig4}. It modulates according to $A\cos[\omega(t-t_{0})]$
with period and phase consistent with 1 year and 2nd June,
respectively. The probability of absence of modulation is
$\sim4\times10^{-4}$. The DAMA global results \cite{Ber99} (NaI,
1, 2, 3, 4 runnings) in the case of assuming the WIMP
interpretation, lead to a WIMP of mass and cross-section given by
$M_{W}=(52^{+10}_{-8})GeV$
$\xi\sigma^{p}=(7.2^{+0.4}_{-0.9})\times10^{-6}pb$. A maximum
likelihood favours the hypothesis of presence of modulation with
the above $M_{W}$, $\xi\sigma_{p}$ values at $4\sigma$ C.L. The
($\sigma$, m) region for spin independent coupled WIMP is the
"triangle" zone depicted in Fig. \ref{fig2}. (We refer to the A.
Incicchitti contribution to these Proceedings and
Ref.\cite{Ber99}). An extension of DAMA up to 250 Kg of NaI
(LIBRA) is being prepared.

The DAMA results have aroused great interest and controversies. It
is imperative to confirm the DAMA results by other independent
experiments with NaI (like LIBRA, NAIAD, ANAIS, or ELEGANTS) and
with other nuclear targets. For instance, CUORICINO, with 42 Kg of
TeO$_{2}$, now being mounted. The DAMA $\sigma$(m) region is being
explored, also by the standard method followed for excluding
WIMPs. Various experiments have already reached and (partially)
excluded the DAMA ($\sigma$, m) region. In particular CDMS,
EDELWEISS and IGEX have derived exclusion contours which "enter"
the DAMA region (see Fig. \ref{fig2}).

The OSAKA group is performing a search with the ELEGANTS V NaI
detector in the underground facility of Oto. ELEGANTS
\cite{elegants} uses huge mass of NaI scintillators (760 Kg)
upgraded from a previous experiment. The background at threshold
is still high. A search for annual modulation did not show any
indication of modulation.

\section{WIMP searches with Xenon Scintillation Detectors}

The search for WIMPs with Xenon scintillators benefits of a
well-known technique. Moreover, background discrimination can be
done better than in NaI.

One of the pioneer searches using Xenon is the DAMA liquid-Xenon
experiment. The spectra of limits on recoils in WIMP-$^{129}$ Xe
elastic scattering using PSD and exclusion plots were published in
Ref.\cite{Ber98}. Recent results of the DAMA liquid Xenon
experiment refers to limits on WIMP-$^{129}$ Xe inelastic
scattering \cite{Ber00}. (See the P.L. Belli contribution to these
Proceedings.)

The ZEPLIN Program \cite{spooner} uses a series of Xenon-based
scintillators devices able to discriminate the background from the
nuclear recoils in liquid or liquid-gas detectors in various ways.
Either using the Scintillation Pulse Shape or measuring the
scintillation and the ionization (an electric field prevents
recombination, the charge being drifted to create a second
scintillation pulse), and capitalizing the fact that the primary
(direct) scintillation pulse and the secondary scintillation pulse
amplitudes differ for electron recoils and nuclear recoils. The
secondary scintillation photons are produced by proportional
scintillation process in liquid-Xenon like in the ZEPLIN-I
detector, (where a discrimination factor of 98\% is achieved) or
by electroluminiscence photons in gas-Xenon (like in the case of
the ZEPLIN-II detector prototype) in which the electrons
(ionization) are drifted to the gas phase where
electroluminiscence takes place (the discrimination factor being
$\mathrm{>99\%}$) \cite{zeplin}. Some prototypes leave been tested
and various different projects of the ZEPLIN series are underway
\cite{cline,spooner} to be installed in Boulby. (See S. Hart
contribution to these Proceedings).
%ZEPLIN-I (UKDMC) is a liquid Xenon 4 Kg detector which use PSD and is running. ZEPLIN-II (UCLA-Torino-Padova-CERN and UKDMC) is a
%1 Kg prototype (Fig. []) used as proof-of-principle of the two-phase (Liquid-Gas) Xenon chamber, which uses as discrimination
%method the measure of  scintillation and luminescence (Fig. []). Extensions up to 30-40 Kg are under way. The ZEPLIN-III (Boulby
%Dark Matter Collaboration: BDMC) is a two phase Xenon device of 6 Kg using ionization and scintillation where a high electric
%field (20KV) has been applied to get low threshold. Finally ZEPLIN IV or ZEPLIN-MAX (BDMC) is a 1 Ton Project of a two-phase
%Xenon chamber which is a modular extension of ZEPLIN-II (Ref. []).

\section{WIMP searches with Time Projection Chambers}

DRIFT is a detector project sensitive to directionality
\cite{drift}. It uses a low pressure (10-40 Torr) TPC with Xenon
to measure the nuclear recoil track in WIMP-Nucleus interactions.
The diffusion constrains the track length observable but DRIFT
reduces the diffusion (transversal and longitudinal) using
negative ions to drift the ionization instead of drift electrons:
gas CS$_{2}$ is added to capture electrons and so CS$^{(-)}_{2}$
ions are drifted to the avalanche regions (where the electrons are
released) for multiwire read-out (no magnetic field needed). The
negative ion TPC has a millimetric diffusion an a millimetric
track resolution. The proof-of-principle has been performed in
mini-DRIFTs, where the direction and orientation of nuclear
recoils have been seen. The event reconstruction, the measurement
of the track length and orientation, the determination of dE/dx
and the ionization measurement permit a powerful background
discrimination (99.9\% gamma rejection and 95\% alpha rejection)
leading to a rate sensitivity of $R<10^{-2(-3)}c/kgday$. DRIFT
will permit to recognize the forward/backward asymmetry and the
nuclear recoils angular distribution, which, as already noted, are
the most clear distinctive signatures of WIMPs. That will permit
hopefully the identification of WIMP. A DRIFT prototype of 1
m$^{3}$ is under construction \cite{spooner}. A project of 10
m$^{3}$ (Xe) scaling up the TPC of 1 m$^{3}$ (Xe) is under way.
(See S. Hart contribution to these Proceedings).

\section{WIMP searches with metastable particle detectors}

WIMP detectors, which use the metastability of the medium where
the nuclear targets are embedded, are the (novel) superheated drop
detectors (SDD) (like SIMPLE and PICASSO) and the (old)
superconducting superheated grains (SSD) (like ORPHEUS). The SDD's
consist of a dispersion of droplets ($\oslash\sim10\mu m$) of
superheated liquid (freon) in a gel matrix. The energy deposition
of a WIMP in the droplets produces a phase transition from the
superheated to normal state causing vaporization of droplets into
bubbles ($\oslash\sim 1mm$), detected acoustically. SSD are
essentially intensitive to low LET particles (e, $\gamma$, $\mu$),
and so good for detecting WIMPs and neutrons. See
Ref.\cite{simple} for SIMPLE and the contribution of V. Zacek to
these Proceedings and \cite{picasso} for PICASSO.

The Superconducting Superheated Grains (SSG) detectors, like
ORPHEUS, is based on the change of phase from the superconducting
superheated to the normal state produced by the WIMP energy
deposition in micrograins inside a magnetic field, at very low
temperatures. The signal is detected through the disappearance of
the Meissner effect. The SSG offer good background rejection
(97\%) (a single grain is expected to flip per WIMP or nucleon
interaction, in contrast to several grains in the case of other
particles), and are sensitive to very low energy deposition (as
proved in neutron irradiation experiments). An experiment with tin
micrograins has just started at the Bern Underground Laboratory
(70 m.w.e.). (See the F. Hasenbalg contribution to these
Proceedings and Ref.\cite{orpheus,pretzl}).

\section{WIMP searches with Cryogenic Particle Detectors}

Thermal detectors \cite{fiorini,pretzl} measure the total recoil
energy transferred by the WIMP to the nucleus in an absorber, via
its small temperature increase (of the order of $\mu K$ if the
working temperature is a few $mK$) with a suitable sensor. They
have quenching factors of about unity ($E_{vis}\sim E_{R}$) and
present other advantages over conventional detectors in the search
for WIMPs: a better energy resolution (much smaller energy quanta
involved in the processes), a very low effective threshold and a
wide choice of absorbers or nuclear targets. Moreover, bolometers
which also collect charge (or light) can simultaneously measure
the phonon and ionization (or scintillation) components of the
energy deposition providing a unique tool of background
substraction and particle identification.

There exists five cryogenic experiments looking for WIMPs
currently running (MIBETA, CRESST, ROSEBUD, CDMS, and EDELWEISS),
another one (CUORICINO) being mounted and a big project (CUORE),
in preparation.
%They can be classified in three classes according to wether they measure
%only phonons, phonons plus charge or phonons plus light. Tables 5, 6and 7 show the main characteristics and performances of the
%cryo-detectors currently in operation.

\begin{figure}[t]
\centerline{\includegraphics[height=5.4cm]{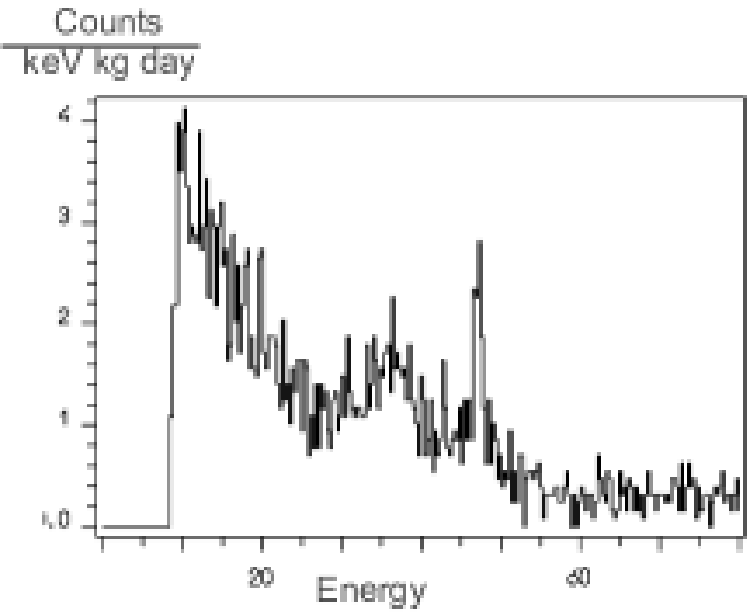}} \caption{}
\label{fig5}
\end{figure}

The MIBETA experiment \cite{mibeta} consists of an array of 20
bolometers of TeO$_{2}$ of 340 g each (the largest cryogenic
experiment now in operation), arranged in a tower-like structure
running in Gran Sasso. The quenching factor (Q) has been measured
to be about 1 (both for Te and O recoils). Although optimized for
double beta decay it has produced interesting results in WIMP
searches. The low energy background collected with two crystals
(M=680 g and t=89 days) with a software threshold of 10 KeV and a
hardware threshold of 2-4 KeV is shown in Fig. \ref{fig5}. The
exclusion plot derived for such background spectrum gives the best
$\sigma$-bound obtained from thermal (only phonons) detectors.
(See A. Giulani contribution to these Proceedings.)

CUORE \cite{cuore} (Cryogenic Underground Observatory for Rare
Events) is an experiment projected as a large extension of MIBETA.
A first step is CUORICINO, a set of 56 crystals of 760 g of
TeO$_{2}$ in a tower-structure similar to MIBETA, which is now
being mounted. Threshold and energy resolution of the 760 g
detectors are $\sim$5 KeV and $\Gamma(46 KeV)=1\pm 0.15$ KeV. In
the case of a (flat) background of $10^{-1}$ \textit{c/KeV Kg
day}, which could be reasonably expected extrapolating the current
MIBETA results, CUORICINO would permit the exploration of the full
DAMA region in 42 Kg year of exposure. (See A. Giulani
contribution to these Proceedings and Ref. \cite{pot}.) The CUORE
project will consist of 1000 detectors (M=760 Kg), forming a cubic
structure to be installed in a special dilution refrigerator in
the Gran Sasso Laboratory.

\begin{figure}[t]
\centerline{\includegraphics[height=5.4cm]{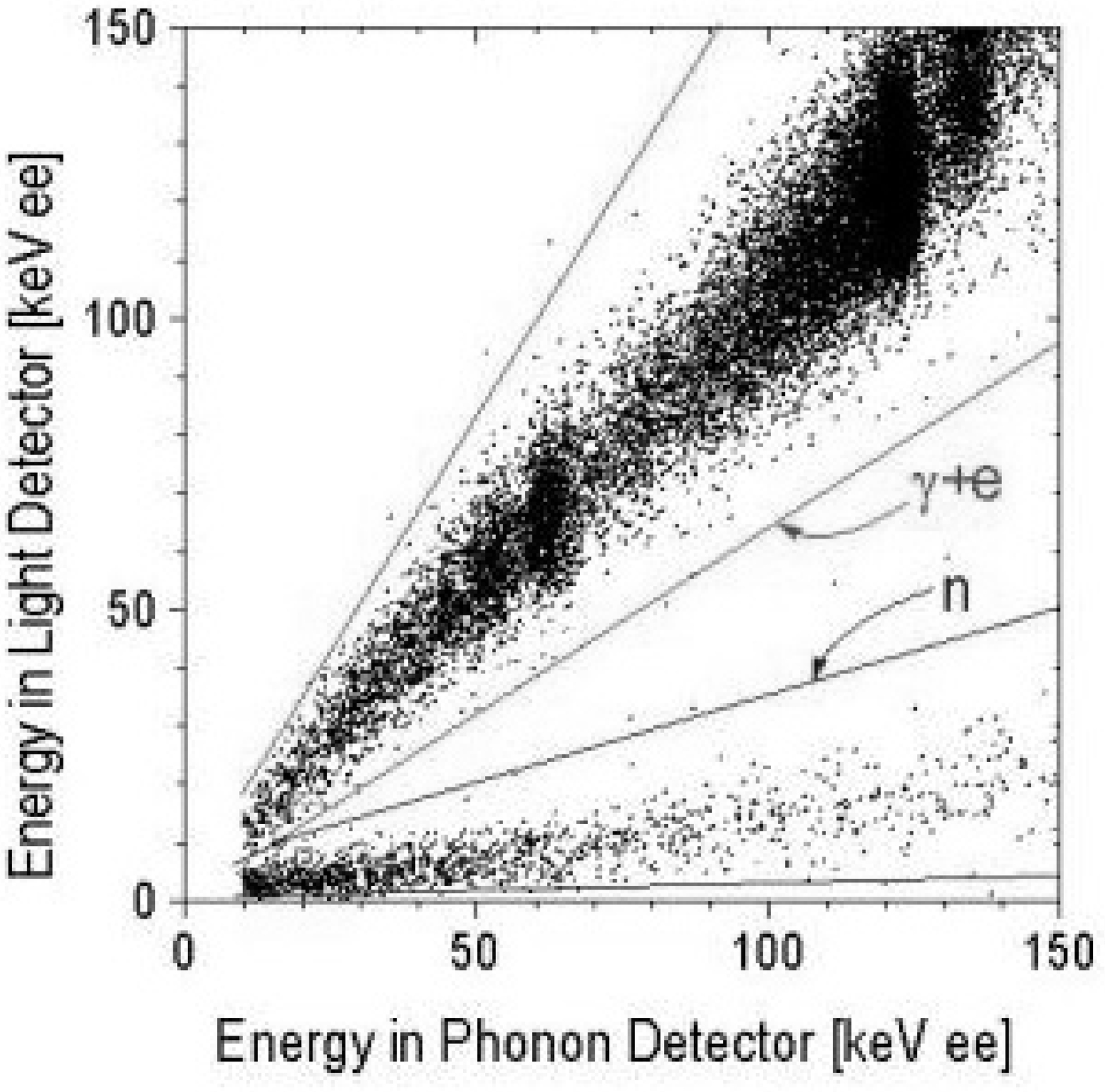}} \caption{}
\label{fig6}
\end{figure}

The CRESST (Cryogenic Rare Event Search with Superconducting
Thermometers) experiment, running also in Gran Sasso, has
initiated a new phase which uses scintillating bolometers of
calcium tungstate ($CaWO_{4}$) to measure simultaneously the heat
and the scintillation produced by the WIMP energy deposit. The
calcium tungstate absorbers of 300 g are viewed by optical
bolometers. Fig. \ref{fig6} shows the discrimination obtained by
the CRESST scintillating bolometer, ranging from 99.5\% to 99.7\%
above $\sim$20 KeV. (See H. Proebst contribution to these
Proceedings.) For the Phase I of CRESST, which used thermal (only
phonons) detectors of 260g of sapphire, we refer to
Ref.\cite{cresst}.

ROSEBUD I (Rare Objects Search with Bolometers Underground) is a
bolometer experiment currently running in Canfranc. The first
phase of the experiment was dedicated to the understanding and
reduction of the radioactive background following successive
removals of the radioimpure materials. Sapphire (25g, 50g) and
germanium (67g) absorbers of excellent energy resolution were used
\cite{rosebud}. Thresholds respectively lower than 1 KeV and 450
eV were achieved on these detectors. Then ROSEBUD has operated
simultaneously bolometers of Ge (67g), $Al_{2}O_{3}$ (50g) and
$CaWO_{4}$ (54 g). For the first time three different absorbers
have been running in the same radioactive environment to help in
understanding the background and hopefully permit to investigate
the nuclear target dependence of the rate.

\begin{figure}[t]
\centerline{\includegraphics[height=5.0cm]{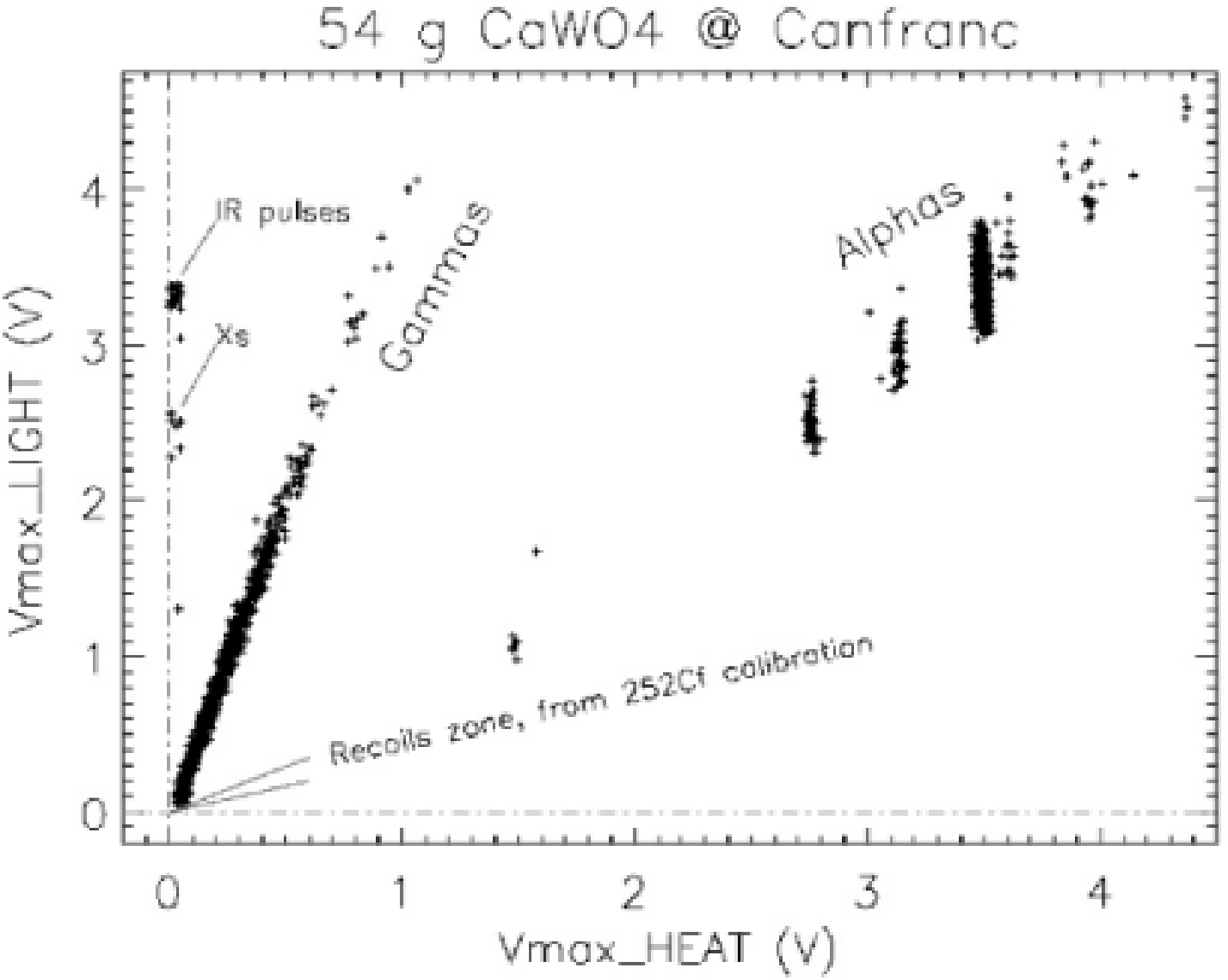}} \caption{}
\label{fig7}
\end{figure}

\begin{figure}[htb]
\centerline{\includegraphics[height=8.4cm]{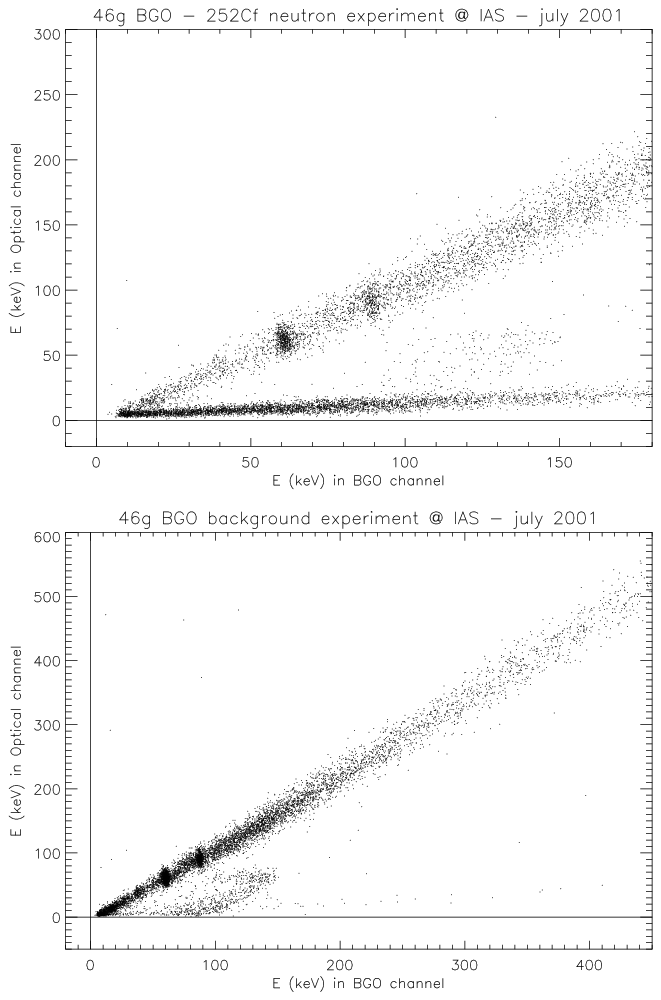}} \caption{}
\label{fig8}
\end{figure}

The new phase of ROSEBUD will use scintillator bolometers of
$CaWO_{4}$ (54 g), and of BGO (46 g), to discriminate backgrounds.
The scintillating bolometers are mounted facing an small optical
bolometer of Ge. Neutron calibrations have shown their high
discrimination power. The CaWO$_{4}$ crystal, tested under neutron
irradiation at Orsay, revealed relative light/heat amplitude
ratios of 10 | 2.5 | 1 for, respectively, gammas | alphas |
recoiling nuclei. Then, in the first underground
heat-scintillation discrimination, in Canfranc, no recoil events
were observed and an alpha contamination was clearly discriminated
from $\gamma$/$\beta$ background. The discrimination plot of Fig.
\ref{fig7} was recorded (15 hours) in a background running in
Canfranc. Threshold was about 45~keV on the heat channel, mainly
due to microphonics. The scintillating bolometer of BGO
(Bi$_{4}$Ge$_{3}$O$_{12}$) (46 g) showed also a high
discrimination power in a $^{252}$Cf neutron irradiation and in a
background running at Canfranc (see Fig. \ref{fig8}). The
light/heat factors and the collected light yields ($\approx$0.8\%
for gammas) are found very similar to those of CaWO$_{4}$, while
the threshold is much lower for the BGO bolometer, bounded to 6
KeV. These features make the BGO bolometers very attractive
candidates for a Dark Matter experiment. (See P. de Marcillac
contribution to these Proceedings.)

CDMS-I (Cryogenic Dark Matter Search) is a cryogenic experiment
installed in the Stanford Underground Facility, which uses Si and
Ge hybrid bolometer-diodes, to discriminate nuclear recoils from
electron recoils by simultaneous measurement of heat and charge.
The rationale is that the ionization Yield (charge/heat), is
different for nuclear recoils (WIMPs, n) than for electron recoils
(produced by the typical background $\gamma$, e, ...). Typical
values of the ionization yield are 1 for electrons bulk events,
0.75 for electrons surface events and 0.3 for nuclear recoil. The
electron-hole pairs are efficiently collected in the bulk of the
detector, but the trapping sites near the detector surface produce
a layer ($10-20 \mu \rm m$) of partial charge collection, where
surface electrons from outside suffer ionization losses and fake
nuclear recoils. This is an important limitation of this
technique, which recently has been addressed and minimized (95\%
discrimination against surface electron background).

\begin{figure}[htb]
\centerline{\includegraphics[height=6.4cm]{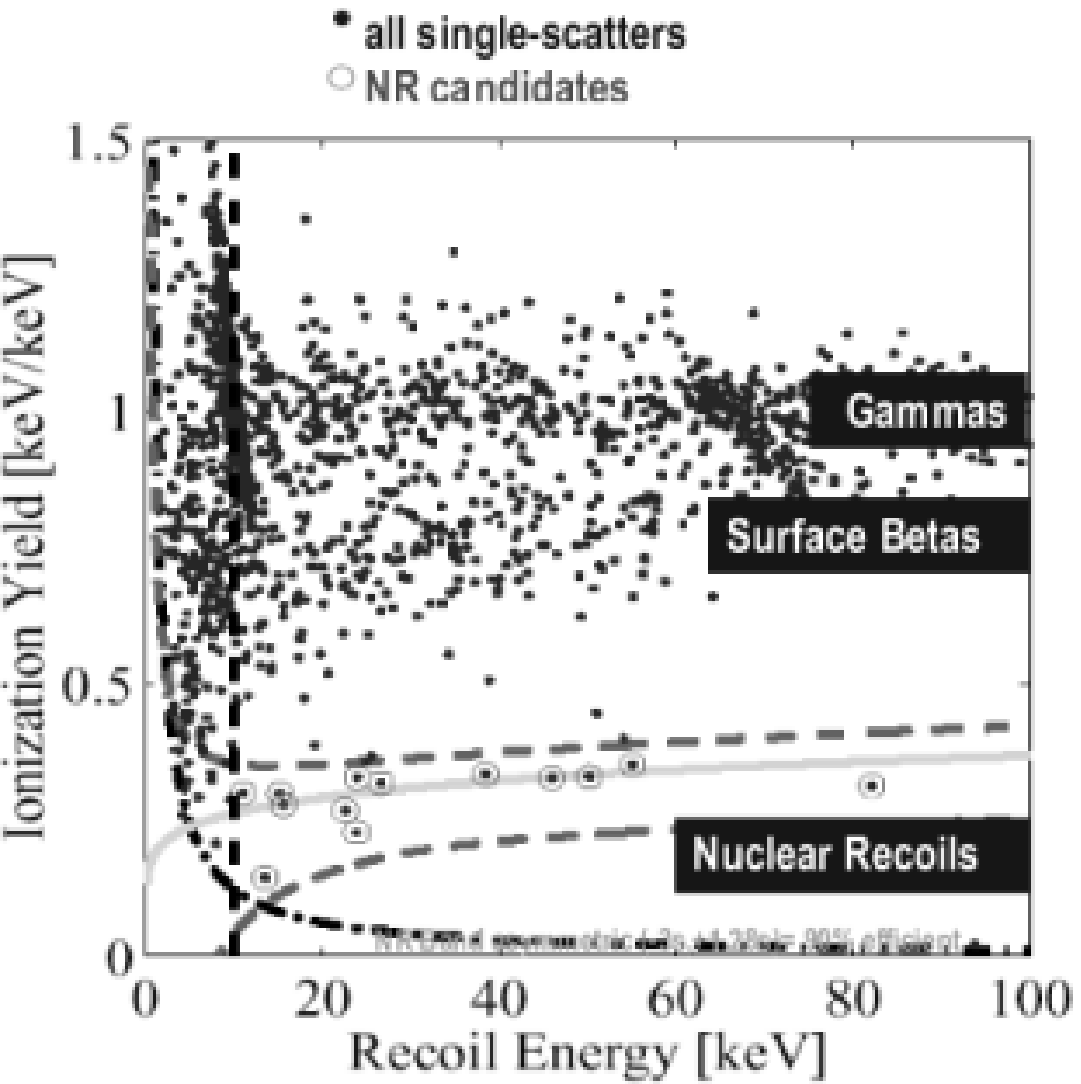}} \caption{}
\label{fig9}
\end{figure}

\begin{figure}[htb]
\centerline{\includegraphics[height=5.4cm]{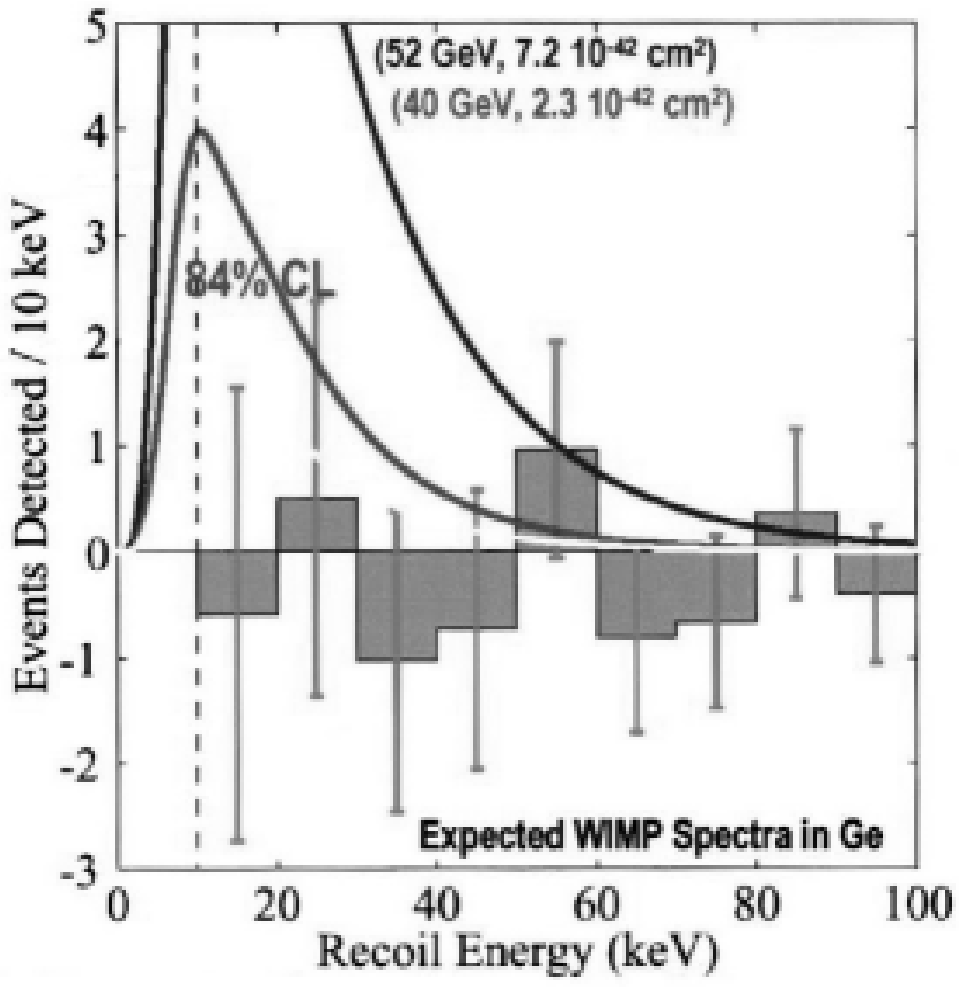}} \caption{}
\label{fig10}
\end{figure}

Two types of detectors and read-out techniques have been
developed: the BLIP germanium detectors where the thermal phonons
are read with NTD (Ge) thermistors (in milliseconds). In the ZIP,
silicon detectors, athermal phonons are detected (microsecond time
scale) with superconducting transition edge thermometers in
tungsten. The total deposited energy is measured calorimetrically.
The charge is measured as in conventional diodes. The main
features and results of CDMS-I have been published in
Ref.\cite{Abusaidi:2000}. An effective exposure of 10.6
\textit{Kgday}, with three BLIP-Ge detectors of 165 \textit{g}
each, have been used for exclusion plots. The almost perfect
discrimination capability of the detector against bulk electron
recoils permits to distinguish the nuclear recoil events. A total
of 13 nuclear recoil events above 10 KeV can be seen
(corresponding to 1.2 events/Kgday, see Fig. \ref{fig9}) and are
due basically to neutrons. The CDMS events after substraction of
the neutron background is shown in Fig. \ref{fig10}, compared with
the expected WIMP spectra in Ge for typical m, $\sigma$ values of
the DAMA WIMP. CDMS and DAMA are incompatible at 99.9\% C.L.
Exclusion plots obtained from the CDMS data are drawn jointly with
that derived from IGEX and EDELWEISS for comparison in Fig.
\ref{fig2}, as all three exclude a fraction of the DAMA region
(see also B. Sadoulet contribution to these Proceedings).

As the neutrons limits the sensitivity of CDMS-I, the next step is
to go underground in the Soudan Mine. The objectives are to reduce
the neutron background from 1 n/Kgday down to
$10^{-2}-10^{-3}$n/Kgday, and to increase substantially the
detector mass up to $\sim 7$ Kg (5 Kg of Ge and 2 Kg of Si) with
Ge of 250 g and Si of 100 g. The goal of CDMS-II is to decrease
the background below $10^{-2}-10^{-3}~c/Kgday$ for $E_{vis}>10KeV$
with the propose of testing S.I. cross-section interactions
neutralino-proton down to $\sim10^{-8}$ pb, for WIMP masses 50-100
GeVs, i.e., well inside the MSSM predictions (see T. Saab
contribution to these Proceedings).

\begin{figure}[t]
\centerline{\includegraphics[height=5.1cm]{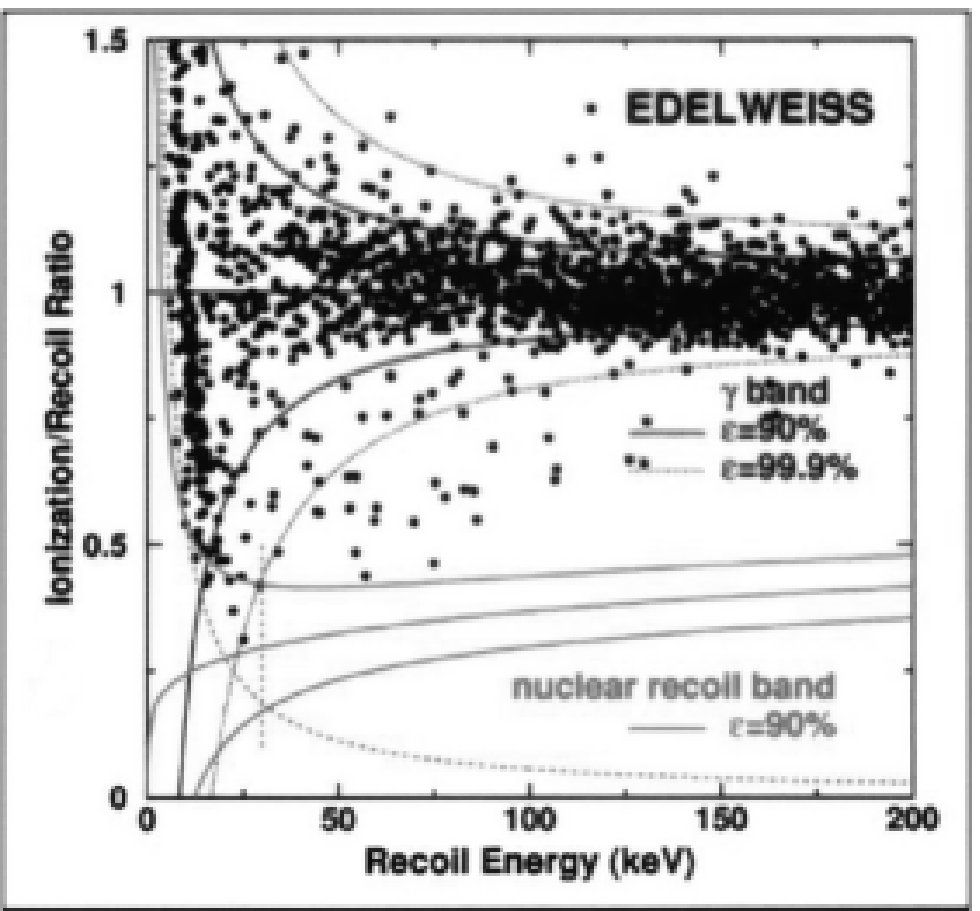}} \caption{}
\label{fig11}
\end{figure}

\begin{figure}[htb]
\centerline{\includegraphics[height=5.4cm]{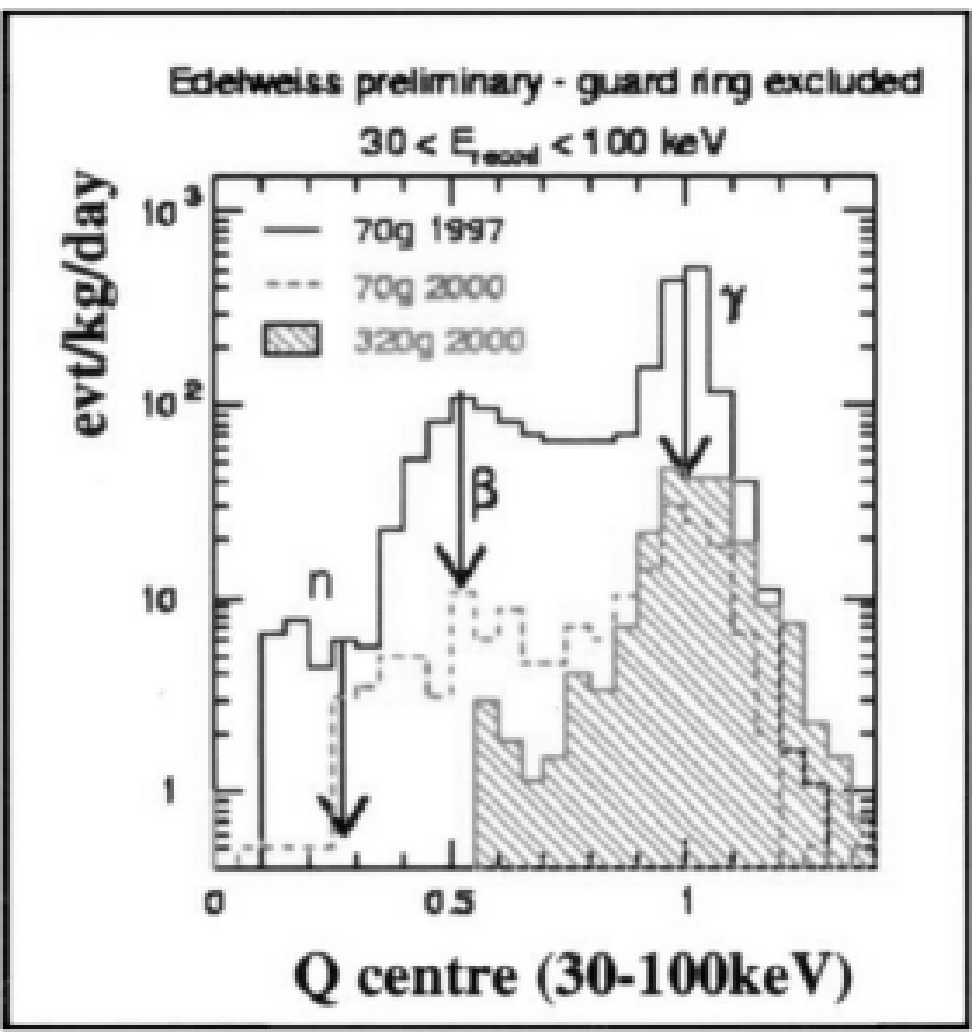}} \caption{}
\label{fig12}
\end{figure}

EDELWEISS is a cryogenic experiment with Ge hybrid detectors which
measure simultaneously ionization and heat \cite{edel70}.
EDELWEISS has operated Ge hybrid bolometers of 70g and 320g in the
Frejus tunnel (LSM) but recent results refer to detectors of 320g
(7cm$\oslash\times$2cm). (For the EDELWEISS phase of 2$\times$70g
Ge detectors see Ref.\cite{edel70}.) The top electrode is divided
in a central disk and a guard ring. The preliminary results (2001)
show a rate B=(20-100)KeV=1.8 \textit{c/KeV Kg day} (with the
central electrode, 54\% volume). After discrimination, the
background turns out to be B$\sim~$0 in 30-100 KeV: no nuclear
recoils were observed in 5.3 Kgday (at the central electrode).
That permits to derive an excellent and reliable (without neutron
substraction) exclusion plot \cite{Benoit:2001} for
spin-independent interaction, which excludes the upper right part
of the DAMA region. Fig. \ref{fig11} shows the physics data
takings with the GeAl6 detector 9.31 Kgday (total exposure with
5.3 Kgday from the center electrode). Fig. \ref{fig12} shows the
successive background spectra obtained by EDELWEISS with the 70 g
and 320 g detector. (See O. Martineau contribution to these
Proceedings.)

The Collaboration plans to install twenty Ge detectors of 320 g
(in an innovative dilution refrigerator) in the next two--three
years, improve the rejection up to 99.99\% and get a background of
$10^{-3}$ c/(keV kg day).

\section{Conclusions and outlook}

The direct search for WIMP dark matter proceeds at full strength.
More than twenty experiments on direct detection illustrate the
effort currently being done. New, dedicated experiments are
focusing now in the identification of WIMPs, discriminating the
nuclear recoils from the background, rather that in constraining
or excluding their parameters space. Their current achievements
and the projections of some of them have been reviewed in this
talk.

\begin{figure}[b]
\centerline{\includegraphics[height=4.5cm]{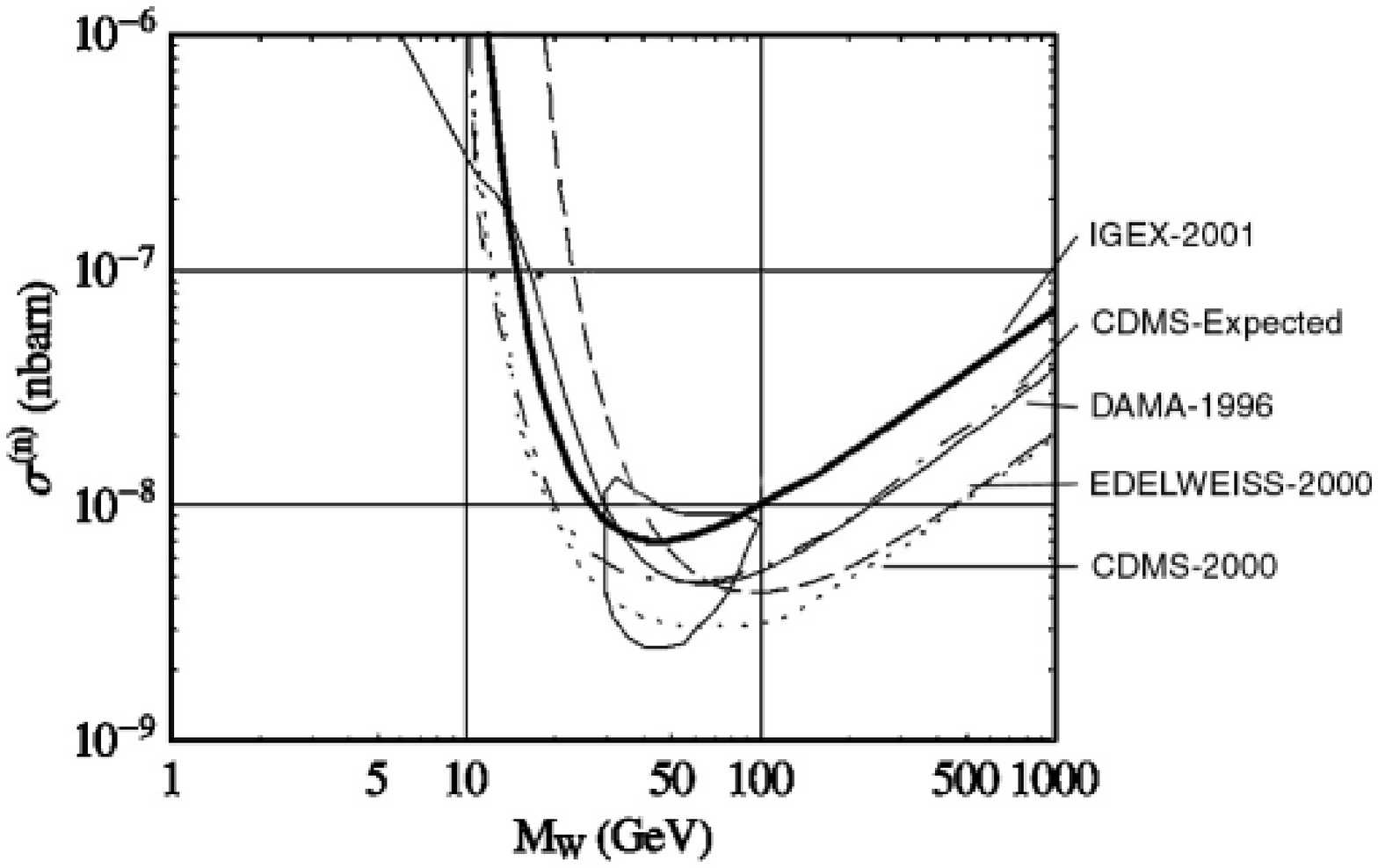}} \caption{}
\label{fig13}
\end{figure}

The present experimental situation can be summarized as follows:
the rates predicted for SUSY-WIMPs extend from 1-10 c/Kgday down
to $10^{-4}-10^{-5}$ c/Kgday, in scatter plots, obtained within
MSSM as basic frame implemented in various alternative schemes. A
small fraction of this window is testable by some of the leading
experiment. The rates experimentally achieved stand around 1
c/Kgday (0.1 c/Kgday at hand) (CDMS, EDELWEISS) and differential
rates $\sim0.1-0.05$ \textit{c/KeV Kg day} have been obtained by
IGEX and H/M, in the relevant low energy regions. The deepest
region of the exclusion plots achieved stands around a few
$\times10^{-6}$pb, for masses 50-200 GeV (DAMA, CDMS, EDELWEISS,
IGEX). The current status of the best exclusion plots is depicted
comparatively in Fig. \ref{fig13}. There exists an unequivocal
annual modulation effect (see Fig. \ref{fig4}) reported by DAMA
(four yearly periods), which has been shown to the compatible
(DAMA) with a neutralino-WIMP, of m$\sim50-60$GeV and
$\sigma^{Si}_{n}\sim 7\times10^{-6}$pb. Recent experiments exclude
at greater or lesser extend (CDMS, EDELWEISS, IGEX) the DAMA
region.

To reach the lowest rates predicted ($10^{-5}$ c/Kgday) in
SUSY-WIMP-nucleus interaction, or in other words, to explore
coherent interaction cross-sections of the order of
$10^{-9}-10^{-10}$pb, substantial improvements have to be
accomplished in pursuing at its best the strategies reviewed in
this talk, with special emphasis in discriminating the type of
events. These strategies must be focussed in getting a  much lower
background (intrinsic, environmental, ...) by improving
radiopurity and shieldings. The nuclear recoil discrimination
efficiency should be optimized going from above 99.7\% up to
99.9\% at the same time that the energy $E_{vis}$ at which
discrimination applies should be lowered. The measurement of the
parameters used to discriminate background from nuclear recoils
should be improved and finally one needs to increase the target
masses and guaranty a superb stability over large exposures. With
these purposes various experiments and a large R+D activity are
under way. Some examples are given in Table \ref{table:1}. The
conclusion is that the search for WIMPs is well focused and should
be further pursued in the quest for their identification.

\begin{table*}[htb]
\caption{WIMP Direct Detection Prospect} \label{table:1}
\begin{tabular}{ll}
\hline
 &BEING INSTALLED/OR PHASE II EXPERIMENTS~(To start
2001-2002)
\\ \hline

CDMS-II & (Ge,Si) Phonons+Ioniz 7 Kg, B$\sim 10^{-2}-10^{-3}$
c/Kgd, $\sigma \sim 10^{-8}$ pb \\

EDELWEISS-II & (Ge) Phonons+Ioniz 6.7 Kg, B$\sim 10^{-2}-10^{-3}$
c/Kgd, $\sigma \sim 10^{-8}$ pb (40-200 GeV) \\

CUORICINO & TeO$_{2}$ Phonons 42 Kg, B$\sim 10^{-2}$ dru, $\sigma
\sim 0^{-7}$ pb
\\

CRESST-II & CaWO$_{4}$ Phonons+light, $B<10^{-2}-10^{-3}$ dru (15
KeV), $\sigma \sim 10^{-7}-10^{-8}$ pb \\ & (50-150 GeV) \\

IGEX &  Ge Ioniz 2.1 Kg, B$< 10^{-1}-10^{-2}$ dru, $\sigma \sim
2\times 10^{-6}$ pb (40-200 GeV) \\

HDMS &  Ge Ioniz 0.2 Kg, $\sigma \sim 6 \times 10^{-6}$ pb (20-80
GeV)
\\

ANAIS & NaI Scintillators 107-150 Kg, B(PSD)$\leq 0.1$ dru,
$\sigma \sim 2 \times 10^{-6}$ pb \\

NAIAD & NaI Scintillators 10-50 Kg, B(PSD)$\leq 0.1$ dru, $\sigma
\sim 10^{-6}$ pb (60-200 GeV) \\ \hline

& IN PREPARATION~(To start 2002-2003)  \\ \hline

LIBRA (DAMA) &  NaI Scintillators 250 Kg \\

GENIUS-TF & Ge Ioiniz 40 Kg, B$<10^{-2}$ dru, $E_{Thr}$=10 KeV
$\rightarrow \sigma \sim 10^{-6}$ pb (40-200 GeV), \\

& $E_{Thr}$=2 KeV $\rightarrow \sigma \sim 10^{-7} pb$ (20-80 GeV)
\\

ZEPLIN-II & Xe-Two-phase 40 Kg, NR discrim$>$99\%, B$<10^{-2}$
dru, $\sigma \sim 10^{-7}$ pb \\

DRIFT-I  &  Xe TPC 1 m$^{3}$, B$<10^{-2}$ dru, $\sigma \sim
10^{-6}$ pb (80-120 GeV) \\ \hline

 & IN PROJECT~($>$2003-2005) \\ \hline

CUORE & TeO$_{2}$ Phonons 760 Kg, $E_{Thr}\sim$2.5 KeV, B$\sim
10^{-2}-10^{-3}$ dru, $\sigma \sim 5\times 10^{-8}$ pb \\

GENIUS 100 & Ge ioniz 100 Kg, $E_{Thr}\sim$10 KeV \\

(GENINO) & B$\sim 10^{-3}-10^{-5}$ dru, $\sigma \sim 5\times
10^{-8}-2\times 10^{-9}$ pb
\\

GEDEON & Ge ioniz 28-112 Kg, B$\sim 2\times 10^{-3}$ dru ($>$10
KeV) $\sigma \sim 10^{-7}-10^{-8}$ pb (40-200 GeV) \\ \hline

& THE FUTURE~($>$2005-2007)  \\ \hline

DRIFT 10 &   Xe 10 m$^{3}$ TPC, $\sigma \sim 10^{-}8$ pb \\

ZEPLIN-MAX & Xe Two-Phase, $\sigma \sim 10^{-10}$ pb \\

GENIUS & Ge ioniz 1-10 Tons, $\sigma \sim 10^{-9}-10^{-10}$ pb
\\

DRIFT-1 ton & Xe 1 Ton TPC, $\sigma \sim 10^{-10}-10^{-11}$ pb
\\ \hline
\end{tabular}\\[2pt]
\end{table*}

\section*{Acknowledgments}

I wish to thank S. Cebri\'{a}n and I.G. Irastorza for their invaluable
collaboration in the making of the exclusion plots and to J.
Morales for useful discussions. The present work was partially
supported by the CICYT and MCyT (Spain) under grant number
AEN99-1033 and by the EU Network contract ERB FMRX-CT98-0167.

\end{document}